\documentclass[prb,nofootinbib,twocolumn,superscriptaddress]{revtex4} 

\usepackage{graphicx}
\usepackage{dcolumn}
\usepackage{bm}
\usepackage{threeparttable}
\usepackage{times}
\usepackage{mathptmx}
\usepackage{lscape}
\usepackage{natbib}
\usepackage{amsmath}
\usepackage{amssymb}
\usepackage{braket}

\def\degree{\kern-.2em\r{}\kern-.3em}

\begin{document}


\title{ Graph Representation for Configurational Properties of Crystalline Solids }

\author{Koretaka Yuge}
\affiliation{
Department of Materials Science and Engineering,  Kyoto University, Sakyo, Kyoto 606-8501, Japan\\
}%

\begin{abstract}
{  We propose representation of configurational physical quantities and microscopic structures for multicomponent system on lattice, by extending a concept of generalized Ising model (GIM) to graph theory. We construct graph Laplacian (and adjacency matrix) composed of symmetry-equivalent neighboring edges, whose landscape of spectrum explicitly represents GIM description of structures as well as low-dimensional topological information in terms of graph. The proposed representation indicates the importance of linear combination of graph to further investigate the role of spatial constraint on equilibrium properties in classical systems. We demonstrate that spectrum for such linear combination of graph can find out additional characteristic microscopic structures compared with GIM-based descriptions for given set of figures on the same low-dimensional configuration space, coming from the proposed representation explicitly having more structural information for e.g., higher-order closed links of selected element. Statistical interdependence for density of microscopic states including graph representation for structures is also examined, which exhibits similar behavior that has been seen for GIM description of the microscopic structures. 
 }
\end{abstract}


\maketitle

\section{Introduction}
In crystalline solids, configurational energetics as well as other physical quantities including elastic modulus, adsorption energy and vibrational free energy can be quantitatively  treated by a generalized Ising model (GIM),\cite{ce0} which can be smoothly combined with {\textit{ab initio}} calculations to predict phase stability and ground-state structures for multicomponent systems.\cite{ce1,ce2,ce3,ce4,ce5}
The GIM can give complete orthonormal basis functions to represent all possible microscopic states on given lattice in configuration space, and enables to completely treat physical quantities as a linear map for the microscopic states. 

Based on GIM, we recently developed a theoretical approach, enabling to characterize physical properties of crystalline solids and their temperature dependences in equilibrium state, by using a few specially-selected microscopic states that are {\textit{independent}} of constituent elements, multibody interactions and temperature.\cite{lsi,emrs} These special states are naturally derived by clarifying how macroscopic physical property connect with spatial constraint on the system. 
These important results rely on the fact that statistical interdependence for density of microscopic states on configuration space become numerically vanished at thermodynamic limit for a wide class of spatial constraint, which has been confirmed based on the theory of Random Matrix.\cite{yuge-RM} 
Our results strongly indicate that information of spatial constraint on the system should be re-examined, especially how the geometrical characteristics of microscopic states on configuration space is restricted by various condition of constraints. 
Our previous studies focus on  microscopic states on configuration space based on GIM descriptions, while other aspect including role of the spatial constraint on low dimensional topology of the structures themselves are not explicitly examined.  Such information can be considered interesting to further address macroscopic properties that do not simply corresponds to linear map with respect to microscopic states on configuration space, including diffusion and deformation behavior for crystalline solids. 

It is thus naturally desired to develop unified representation to simultaneously treat GIM and topological description of structures for further studying the relationship between spatial constraint and macroscopic properties. 
Based on the graph theory, the present study proposes construction of graph Laplacians (and their adjacency matrices) for configurational properties of crystalline solids by a natural extension of GIM, where landscape of their spectrum explicitly contains both GIM and topological description of the microscopic structures: Crystal structures can be considered as linear combination of multiple graph spectrum.
Based on the proposed representation, we demonstrate practical examples of finding characteristic microscopic structures beyond GIM description, and of  
addressing statistical interdependence for microscopic structures on configuration space.


\section{Construction and Concept}
Let us first consider a given lattice $Q$, composed of $N$ lattice points indexed by $i=1,\ldots, N$. 
GIM typically employs complete orthonormal (but orthonormality is not always required) basis functions (CONS) of vector space $V_Q$ for the system, $\psi$s, by taking a tensor product of vector space on individual lattice points, which is generally constructed by applying Gram-Schmidt technique to linearly-independent polynomial set of spin variables (specifying occupation of constituents at lattice point $i$) for $R$-component system, $\left\{1, \sigma_i , \ldots, \sigma_i^{\left(R-1\right)}\right\}$: 
\begin{eqnarray}
V_{Q} &=& \bigotimes_{i=1}^{N} V_{i} \nonumber \\
V_i &=& \mathrm{span}\left\{\phi_0\left(\sigma_i\right),\ldots , \phi_{R-1}\left(\sigma_i\right)\right\},
\end{eqnarray}
where
\begin{eqnarray}
\phi_m\left(\sigma_i\right) &=& f_m\left(\sigma_i\right)/\Braket{f_m\left(\sigma_i\right)|f_m\left(\sigma_i\right)}^{1/2} \nonumber \\
f_m\left(\sigma_i\right)&=&\sigma_i^m -\sum_{j=0}^{m-1}\Braket{\phi_j\left(\sigma_i\right)|\sigma_i^m}\phi_j\left(\sigma_i\right) \quad \left(m\neq 0\right) \nonumber \\
f_0\left(\sigma_i\right) &=& 1. 
\end{eqnarray}
Here, $\Braket{\cdot|\cdot}$ denotes inner product on a given single lattice point considered. 
It is clear from the above equations that CONS of $\Psi$s should have index of both specifying a set of lattice points (i.e., "figure") whose basis function is not  unity, $l$, and a set of index of basis function (i.e., subscript of $\phi$), $\alpha$: $\psi_l^{\left(\alpha\right)}$. 
With these definition, any physical quantity $J$ as a linear map for microscopic states on lattice, $\vec{\sigma}$, can always be given by
\begin{eqnarray}
\label{eq:ce}
J\left(\vec{\sigma}\right) = \sum_{l\in \textrm{figure}}\sum_{\left(\alpha\right)} \Braket{J|\psi_{l}^{\left(\alpha\right)}} \psi_{l}^{\left(\alpha\right)}\left(\vec{\sigma}\right), 
\end{eqnarray}
where $\Braket{\cdot|\cdot}$ denotes inner product over whole lattice points in $Q$.

In order to include the information of basis functions for GIM in graph theory, we consider a given lattice $Q$ as a set of graph $G_R=\left(V, E_R\right)$ composed of edges for $R$-th symmetry-equivalent pair figure and their constituent vertices. 
Then we define corresponding graph Laplacian $\bm{L}_R^{\left(\alpha\right)}$ as the sum of three contributions, namely, 
\begin{eqnarray}
\label{eq:lp1}
\bm{L}_R^{\left(\alpha\right)} = \bm{D}_R^{\left(\alpha\right)} - \bm{B}_R^{\left(\alpha\right)} - {}^t\!\bm{B}_R^{\left(\alpha\right)},
\end{eqnarray}
where 
\begin{eqnarray}
\label{eq:lp2}
\bm{B}_R^{\left(\alpha\right)}\left(i,j\right) = \begin{cases} 
\sqrt{\phi_p\left(\sigma_i\right)\phi_q\left(\sigma_j\right)} & \left(p,q \in \alpha, i,j\in R, i < j\right)  \\
0 & (otherwise)
\end{cases}
\end{eqnarray}
and $\bm{D}_R^{\left(\alpha\right)}$ is the diagonal matrix, whose diagonal element $d_{ii}$ is given by
\begin{eqnarray}
\label{eq:lp3}
d_{ii} = \sum_{k\neq i} \bm{B}_R^{\left(\alpha\right)}\left(i,k\right) + {}^t\!\bm{B}_R^{\left(\alpha\right)}\left(i,k\right). 
\end{eqnarray}
Note that index set, $\left(\alpha\right)$, includes "order" of index: e.g., when $\left(\alpha\right)=\left\{p,q\right\}$, then $\left(\alpha\right)\neq\left\{q,p\right\}$.
The sum of $\bm{B} + {}^t\!\bm{B}$ is known as adjacent matrix, $\bm{A}$. 
Here we consider graph Laplacian as symmetric matrix, indicating that whole lattice points on "empty" lattice are symmetry-equivalent, which holds for representative lattice including fcc, bcc, hcp and diamond. 
For simplicity (without lack of generality), hereinafter index $R$ includes both figure type $R$ and set of basis index $\left(\alpha\right)$.  

From the above definitions, we can construct the following relationship between graph Laplacian and basis functions in GIM:
\begin{eqnarray}
\label{eq:gl}
\Braket{\psi \left(\vec{\sigma}\right)}_R = \rho_R^{-1} \sum_{m} C_m\cdot \textrm{Tr} \left[ \left(\sum _{l\in R} \bm{B}_l\left(\vec{\sigma}\right)\pm {}^t\!\bm{B}_l\left(\vec{\sigma}\right) \right)^{N_R} \right]
\end{eqnarray}
when we explicitly consider $\psi$s as \textit{extensive}. 
Here, $C_m\in\mathbb{Z}$ is the coefficient of the following trace, $l$ denotes possible pair subfigure for figure $R$ and $N_R$ is the dimension of figure $R$. $\Braket{\quad}_R$ denotes linear average of basis function for figures with the same dimension as $R$, composed of the same set of pair subfigures for path with $N_R$ walks, and $\rho_R$ denotes number of the possible path considered.
Eq.~(\ref{eq:gl}) indicates that when a given figure $R$ can be specified by a set of all possible pair subfigures, GIM basis functions can be represented by a linear combination of trace of multiple power of upper and lower triangular part of graph Laplacian defined by Eqs.~(\ref{eq:lp1})-(\ref{eq:lp3}). 

\section{Application and Discussions}
\subsection{Graph representation for configurational properties}
We here see the concrete relationship between GIM and graph descriptions in Eq.~(\ref{eq:gl}) with examples of pair, triplet and quartet correlations. 
\subsubsection{$N_R=2$ (Pair Correlation)}
Let us consider a pair correlation for figure $R$, and we define matrix $\bm{A}_R = \bm{B}_R + {}^t\!\bm{B}_R$. From Eqs.~(\ref{eq:lp1})-(\ref{eq:lp3}), we can directly express
\begin{eqnarray}
\mathrm{Tr}\left[\bm{A}_R^2 \right] &=& \sum_i \sum_j \Braket{i|A_R|j}\Braket{j|A_R|i} \nonumber \\
&=& \sum_i \sum_j \phi_p\left(\sigma_i\right)\phi_q\left(\sigma_j\right).
\end{eqnarray}
The last equation can be obtained since diagonal elements of $\bm{A}_R$ are all zero. Therefore, we get
\begin{eqnarray}
\psi_R\left(\vec{\sigma}\right) = \frac{1}{2} \mathrm{Tr}\left[\bm{A}_R^2 \right].
\end{eqnarray}

\subsubsection{$N_R=3$ (Triplet Correlation)}
For triplet correlation, expression of Eq.~(\ref{eq:gl}) can be respectively provided for a set of constituent pair figures. First, we consider the triplet figure $R$, composed of three equivalent pair figures of $r$. 
In this case, number of possible path to construct $R$ is six, i.e., $\rho_R =6$. Similar to the pair correlation, we can immediately get
\begin{eqnarray}
\psi_R\left(\vec{\sigma}\right) = \psi_{\left(rrr\right)}\left(\vec{\sigma}\right)  = \frac{1}{6} \mathrm{Tr}\left[ \bm{A}_r^3\right]
\end{eqnarray}

Next, we consider a triplet, composed of two different pairs of $r$ and $s$. In this case, we consider a linear combination of trace, for instance,
\begin{eqnarray}
&&\mathrm{Tr}\left[ \left(\bm{A}_r + \bm{A}_s\right)^3 \right] - \mathrm{Tr}\left[ \left(\bm{A}_r - \bm{A}_s\right)^3 \right] \nonumber \\
&&= 2\mathrm{Tr}\left[\bm{A}_s^3 \right]  + 2 \sum_{i,j,k\in\left(rrs\right)} \Braket{i|A_{rs}|j}\Braket{j|A_{rs}|k}\Braket{k|A_{rs}|i}, \nonumber \\
\quad
\end{eqnarray}
where
\begin{eqnarray}
\bm{A}_{rs} = \bm{A}_r + \bm{A}_s. \nonumber
\end{eqnarray}
Here, summation for right-hand side is taken over all triplets composed of two pairs of $r$ and one pair of $s$. Therefore, we get for triplet of $R=\left(rrs\right)$ as
\begin{eqnarray}
\psi_{R}\left(\vec{\sigma}\right) = \frac{1}{12}\left\{ \mathrm{Tr}\left[ \left(\bm{A}_r + \bm{A}_s\right)^3 \right] -\mathrm{Tr}\left[ \left(\bm{A}_r - \bm{A}_s\right)^3 \right] -2\mathrm{Tr}\left[ \bm{A}_s^3 \right]      \right\}.  \nonumber \\
\quad
\end{eqnarray} 

Finally, we consider a triplet composed of three different pairs of $r$, $s$ and $t$. Using the characteristic that $\mathrm{Tr}\left[\left(\bm{A}_r + \bm{A}_s + \bm{A}_t\right)^3  \right]$ includes all possible path for third order moment composed of $r,s,t$ pairs, we can get
\begin{widetext}
\begin{eqnarray}
\psi_R\left(\vec{\sigma}\right) =  \frac{1}{6}\left\{ \mathrm{Tr}\left[ \left(\bm{A}_r + \bm{A}_s + \bm{A}_t\right)^3 \right] + \sum_{I=r,s,t} \mathrm{Tr}\left[ \bm{A}_I^3\right] - \sum_{\substack{I, J=r,s,t \\ I \neq J}} \mathrm{Tr} \left[  \left(\bm{A}_I + \bm{A}_J\right)^3 \right] \right \}
\end{eqnarray}
\end{widetext}

As seen, using the proposed approach, we can determine all triplet correlations from linear combination of graph spectrum composed of constituent pair subfigures, which cannot be achieved by conventional GIM description: This is because GIM pair basis functions just contain information about trace of $\bm{A}^2$ shown above.

\subsubsection{$N_R=4$ (Quartet Correlation)}
\begin{figure}[h]
\begin{center}
\includegraphics[width=0.68\linewidth]
{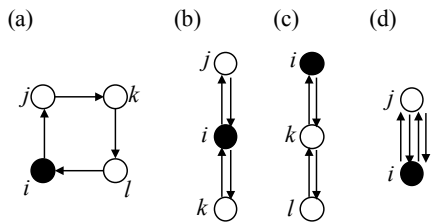}
\caption{Schematic illustration of possible pathways contributing to fourth order moment starting from lattice point $i$. }
\label{fig:4mom}
\end{center}
\end{figure}

When the dimension of figure, $N_R$, is four or larger, the situation is different from $N_R \le 3$. 
Since $N_R$ ($\ge 4$) walk does not always results in figure with dimension of $N_R$, such undesired path should be omitted. Figure~\ref{fig:4mom} shows schematic illustration of possible path contributing to fourth order moment starting from a lattice point  $i$. Clearly, path of (b)-(d) should be omitted, and path of only (a) should be included to obtain quartet correlation. 
Generally, when dimension of a given figure is smaller than number of walk, corresponding path always includes at least a single round trip between certain two lattice points. 
Based on this characteristics, we can vanish undesired path to obtain multisite correlation using linear combination of $\bm{B}$ and ${}^t\!\bm{B}$, i.e., $\bm{B}\pm {}^t\!\bm{B}$. 
For instance, quartet correlation consisting of four neighboring ($r$) and two next neighboring ($s$) pairs shown in Fig.~\ref{fig:4mom} can be given by
\begin{eqnarray}
\Braket{\psi\left(\vec{\sigma}\right)}_{R} = \frac{1}{8} \left\{\mathrm{Tr}\left[ \left(\bm{A}_r + \bm{A}_s\right)^4 \right] - \mathrm{Tr}\left[ \left(\bm{A}'_{r} + \bm{A}'_s\right)^4 \right]    \right\}, \nonumber \\
\quad
\end{eqnarray}
where
\begin{eqnarray}
\bm{A}'_r = \bm{B}_r - {}^t\!\bm{B}_r.
\end{eqnarray}

From above discussions, in addition to structural information from GIM description, spectrum for linear combinations of graph such as $\bm{A}_r \pm \bm{A}_s$ should be focused on, to further investigate the relationship between spatial constraint and microscopic structures: 
They are discussed in the followings.


\subsection{Applications to lattice}

\begin{figure}[h]
\begin{center}
\includegraphics[width=1.04\linewidth]
{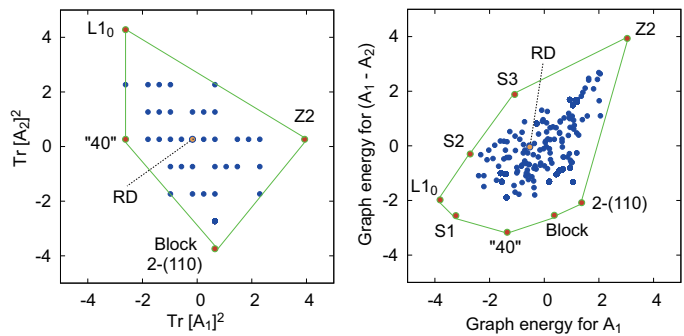}
\caption{(Color online) Left: Constructed configurational polyhedron (CP) for $\mathrm{Tr}\left[\bm{A}_R^2 \right]$ for equiatomic fcc lattice, corresponding to conventional CP based on GIM description. Right: CP in terms of graph energy for a single ($\bm{A}_1$) and linear combination ($\bm{A}_1 - \bm{A}_2$) of graph. Trace and graph energy are standardized based on their linear average and standard deviation. }
\label{fig:cp}
\end{center}
\end{figure}

\begin{figure}
\begin{center}
\includegraphics[width=0.92\linewidth]
{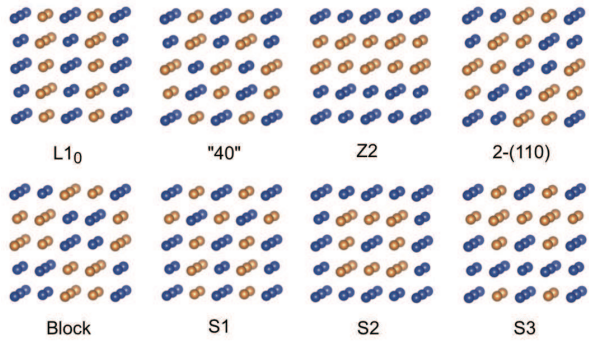}
\caption{(Color online) Atomic configuration on $2\times 2\times 2$ fcc conventional cell at vertices of CP in right-hand side of Fig.~\ref{fig:cp}. }
\label{fig:str}
\end{center}
\end{figure}

While the above discussions for representation of configurational properties based on graph theory can be applied for any choice of basis functions in GIM, we would first construct GIM basis functions so that the resultant Laplacian (and adjacency matrix) form can be well-transferable to existing graph theory for practical use. First, since graph Laplacian is typically a real matrix, 
GIM basis functions should be non-negative real number. This can be easily achieved for multicomponent (including binary) system, where we employ, for instance, non-orthogonal (but complete) basis functions of $\left\{1,\sigma_i, \sigma_i^2,\ldots \right\}$ at each lattice point with non-negative spin variables: These definitions always result in non-negative value of GIM basis functions, which is a desired property. Another notation here is that off-diagonal element of adjacent matrix (or Laplacian) typically takes binary value of 0 or 1 (-1), in which the theoretical studies have actively been performed so far. This can also be easily achieved by the present approach for binary system, where we define GIM basis function on {\textit{individual}} lattice point as $\left\{1,\sigma_i\right\}$ and define spin variables taking 0 and 1: This satisfies that corresponding off-diagonal elements for adjacent matrix (graph Laplacian) takes desired binary values. Simple example of the resultant adjacency matrix in binary system is given in Appendix.  The proposed representation to relate GIM basis functions with graph spectrum shown above strongly indicates that again, not only spectrum for a single graph, but also landscape of spectrum for linear combinations of multiple graphs should be of great importance.

\begin{figure}
\begin{center}
\includegraphics[width=0.77\linewidth]
{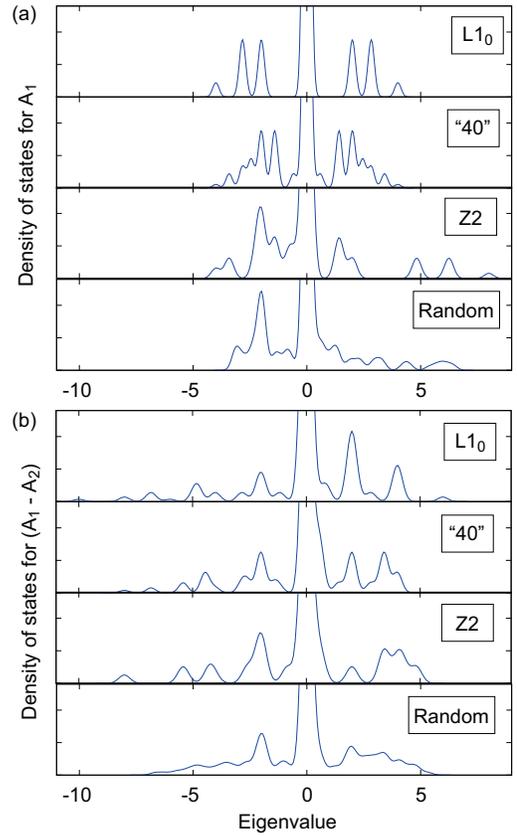}
\caption{(Color online) Density of states for $\bm{A}_1$ (a) and $\left(\bm{A}_1 - \bm{A}_2\right)$ (b) for three ordered structures at the vertices of CP shown in Fig.~\ref{fig:cp} and for quasi-random structure. }
\label{fig:dos1}
\end{center}
\end{figure}

One of such important applications is to examine landscape of the so-called {\textit{configurational polyhedra}} (CP),\cite{cp} which is a hyperpolyhedra in multidimensional configuration space determining upper and lower value for GIM basis functions constrained by a given lattice (or more generally, spatial constraint). 
In classical systems where energy (or other physical quantities such as elastic constants) is given by Eq.~(\ref{eq:ce}), structures that have maximum or minimum physical quantities should always be restricted to those at vertices of CP.  Since the CP is constructed based on non-interacting system, landscape of CP purely depends on the class of spatial constraint (e.g., lattice for crystalline solids). This means that when spatial constraint is once given, we can a priori know a set of "candidate structure" to exhibit extreme physical quantities without any information about energy. Using this characteristics, for instance, nature's missing ordered structures in alloys has been successfully explained\cite{gus} based on low-dimensional figures on lattice. Up to date, exact landscape of CP is not well clarified even for binary system on lattice, so we should numerically construct the CP by directly consider all possible microscopic states within finite system size. Here, the problem is that number of vertices of CP exponentially increases at high-dimensional configuration space considered: It is thus fundamentally important to find out many characteristic structures at low-dimensional space based on geometrically low-dimensional figures, such as symmetry-nonequivalent pairs on lattice. However, it has been shown\cite{cp} that structures at vertices of CP for given configuration space is invariant with linear transformation of the coordination: This means that when we further find out such characteristic structures based on GIM descriptions, we should generally include information about higher-dimensional figures on lattice, but the resultant CP projected onto low-dimensional configuration space typically lose information about vertices that are originally found at low-dimensional space based on low-dimensional figures.\cite{yuge-cp} In order to overcome this problem in GIM, our strategy is thus to construct graph representation for crystalline solids based on constituent pair figures on lattice, whose spectrum not only contains information about GIM pair correlations, but also includes higher-dimensional figures (or links) consisting of the corresponding pairs, which is given by Eq.~(\ref{eq:gl}). The details are shown in the followings.

With these considerations, we first apply the present representation to investigating landscape of CP  in terms of the graph spectrum, which leads to further addressing role of the spatial constraint. Since again, microscopic structure at vertices of CP is known to be invariant by linear transformation of GIM basis functions, constructing CP-like hyperpolyhedron by graph spectrum other than GIM information can be of great interest. 
In the present study, we therefore focus on not only structure information based on GIM, but also the sum of absolute of all eigenvalues for graph spectrum, called "graph energy", which has been extensively investigated to characterize such as regularity by determining upper and/or lower bound of its value, and corresponding application has been done for molecules to relate to its energetics.\cite{ge1,ge2,ge3,ge4} 
We here take examples of all possible atomic configurations on $4\times 4\times 4$ expansion of fcc conventional unit cell having minimal unit consisting of up 16 atoms, and calculate corresponding graph spectrum. 
Figure~\ref{fig:cp} shows the resultant CP in terms of (i) $\mathrm{Tr}\left[ \bm{A}_r^2\right]$ for $r=1$ and 2 ($r$ denotes $r$-th neighboring coordination), and (ii) graph energy for $\bm{A}_1$ and $\bm{A}_1 - \bm{A}_2$: In the figure, "RD" denotes atomic configuration closest to its center of gravity for left-hand CP, and atomic configurations at the vertices of CP are shown in Fig.~\ref{fig:str}. 
The left-hand of Fig.~\ref{fig:cp} corresponds to the standard CP in terms of GIM basis function, where we can successfully find two ordered structures at the vertices of L1$_0$ and "40" that are known to be ground-state atomic configurations for real alloys (e.g., L1$_0$ for CuAu and "40" for PtRh).\cite{ord1, ord2} We also find Z2 structure at the vertex, considered as alternate two-layer stacking along (001) (which is predicted as ground-state for PtRu based on systematic first-principles study\cite{ord3}), and find two additional structures which we call here "Block" and "2-(110)". 
When we see the right-hand of Fig.~\ref{fig:cp}, several characteristics are found: (i) Five atomic configurations at the vertices in left-hand figure are also located at vertices of CP for right-hand figure in terms of graph energy, (ii) structures of "Block" and "2-(110)", which are not distinguished in left-hand figure, are distinguished for CP of graph energy, (iii) three additional characteristic atomic configurations of S1, S2 and S3 at vertices appears, which cannot be achieved by linear transformation of considered GIM basis functions, and (iv) atomic configuration near the center of gravity in the left-hand figure, "RD", which corresponds to quasi-random structure in terms of pair correlations, also locates near the center of gravity for CP of graph energy. Note that "RD" ideally should place at the origin $\left(0,0\right)$ of the CP, since it has no pair correlations. In the present case, since we consider the limited number of atoms, we here define the quasi-random structure closest to the origin. 
Therefore, the new CP based on the proposed representation not only retains characteristic vertices structures found for original CP by GIM description (left-hand of Fig.\ref{fig:cp}), but also finds out  characteristic structures reflecting the information about correlations (or links) for higher-dimensional figures at two-dimensional space, which is the desired property described above: Significant advantage of the proposed representation is demonstrated.  
In addition, landscape of the CP, indicated by green solid lines, certainly reflects the class of spatial constraint on the system where near their boundaries, statistical interdependence of density of microscopic states cannot be neglected:\cite{lsi, emrs} Landscape of CP for right-hand figure therefore should contain information about another aspect of spatial constraint on the system, which cannot be explicitly obtained by GIM description. 
We should note here that when we simply construct CP in terms of graph energy of a single graph (i.e., graph energy in terms of $\bm{A}_1$ and $ \bm{A}_2$), the resultant CP does not satisfy the above characteristics of (i) or (iii), demonstrating that not only a single graph, graph spectrum for their linear combination including negative sum (e.g., $\bm{A}_1 - \bm{A}_2$ as shown in Fig.~\ref{fig:cp}) can be  important to characterize atomic configurations on given lattice, reflecting the role of spatial constraint. 

To further investigate the proposed graph approach shown in Fig.~\ref{fig:cp}, we show in Fig.~\ref{fig:dos1} density of states (DOS) for eigenvalues of $\bm{A}_1$ and $\bm{A}_1 - \bm{A}_2$ for three ordered structures and one quasi-random structure, which are used to constructing CP in Fig.~\ref{fig:cp}. From Fig.~\ref{fig:dos1} (a), we can see that DOSs for ground-state structures of L1$_0$ and "40" are symmetric with respect to zero eigenvalue, while Z2 and random structure are asymmetric. Within the symmetric DOS of L1$_0$ and "40", although their second-order moment exactly takes the same value, their landscape shows clear difference. 
When we consider linear combination, e.g., $\bm{A}_1 - \bm{A}_2$ shown in Fig.~\ref{fig:dos1} (b), (i) such symmetric features of DOS for $\bm{A}_1$ disappears both for L1$_0$ and "40". 
To more quantitatively see the relationship between landscape of the DOSs and structure, we here focus on the differences in symmetry of DOS between $\bm{A}_1$ and $\bm{A}_1 - \bm{A}_2$. 
Since by definition, $\mathrm{Tr}\left[ \bm{A}_R\right]=0$ for all possible pair figure $R$, asymmetry of DOS certainly reflects the third-order moment of its spectrum 
 $\mu_3 \left[ \mathrm{Spec}\left(\bm{A}_R\right)  \right]$, generally called skewness. When we define $\bm{A}_{12}=\bm{A}_1 + \bm{A}_2$ and $\bm{A}_{1\overline{2}}=\bm{A}_1 - \bm{A}_2$, we can quantitatively give the relationship for asymmetry as
\begin{widetext}
\begin{eqnarray}
\label{eq:3m}
&&\mu_3 \left[ \mathrm{Spec}\left(\bm{A}_1 - \bm{A}_2\right)  \right] = N^{-1}\sum_i\sum_j\sum_k \Braket{i|A_{1\overline{2}}|j} \Braket{j|A_{1\overline{2}}|k} \Braket{k|A_{1\overline{2}}|i} \nonumber \\
&&= \mu_3 \left[ \mathrm{Spec}\left(\bm{A}_1 \right)  \right]  - \mu_3 \left[ \mathrm{Spec}\left( \bm{A}_2\right)  \right]  + N^{-1}\left\{\sum_{i,j,k\in \left(122\right)} \Braket{i|A_{12}|j} \Braket{j|A_{12}|k} \Braket{k|A_{12}|i}  - \sum_{i,j,k\in \left(112\right)} \Braket{i|A_{12}|j} \Braket{j|A_{12}|k} \Braket{k|A_{12}|i} \right\}\nonumber \\
&&= \mu_3 \left[ \mathrm{Spec}\left(\bm{A}_1 \right)  \right] - N^{-1}\sum_{i,j,k\in \left(112\right)} \Braket{i|A_{12}|j} \Braket{j|A_{12}|k} \Braket{k|A_{12}|i}.
\end{eqnarray}
\end{widetext}
To derive the last equation from the second one, we employ the geometrical characteristics of triplet figure on fcc lattice. 
From Eq.~(\ref{eq:3m}), it is now clear that difference in asymmetry between $\bm{A}_1$ and $\bm{A}_1 - \bm{A}_2$ certainly reflects the number of closed triplet links consisting of two 1NN pairs  and one 2NN pair figure containing element only of $\sigma=+1$, since the number is proportional to the second term of the last equation. For instance, symmetric landscape of $\mathrm{Spec}\left(\bm{A}_1 \right)$ found for L1$_0$ and "40" means that their is no closed triplet link consisting of three 1NN pairs (i.e., $\left(111\right)$) with element of $\sigma=+1$, while Z2 has strongest asymmetry (i.e., having largest $\mu_3 \left[ \mathrm{Spec}\left(\bm{A}_1\right)  \right]$ among four structures in Fig.~\ref{fig:dos1}) means the largest number of the closed triplet links, and random structure has intermediate $\mu_3$, corresponding to intermediate number of the links. 
Meanwhile, stronger asymmetry of $\mathrm{Spec}\left(\bm{A}_1 - \bm{A}_2 \right)$ for L1$_0$ than that for "40" means the larger number of closed triplet links $\left(112\right)$ with element of $\sigma=+1$ in L1$_0$ than that in "40". 
Such information about multiple links composed of selected element are considered of great significance to consider, for instance, diffusion path of interstitial atoms in crystalline solids.\cite{toyo}
As seen, basically the higher-order moment of $\textrm{Spec}\left(\sum_R \pm A_R\right)$ explicitly contains information about the number of corresponding closed links composed of a set of $\left\{R\right\}$ (note: 2nd-order moment corresponds to information about GIM pair descriptions), resulting in the differences of landscape of DOSs in Fig.~\ref{fig:dos1} as well as in finding additional characteristic ordered structures of S1, S2 and S3, and distinguishment of block and 2-(110) structure in the right-hand side of Fig.~\ref{fig:cp}, while the GIM description with the same set of figure $\left\{R\right\}$ (and their linear combination) again does not explicitly contain such higher-order structural information. 
For instance, although it can be seen Fig.~\ref{fig:dos1} that $\mu_2 \left[ \mathrm{Spec}\left(\bm{A}_1 \right)  \right]$ and $\mu_3 \left[ \mathrm{Spec}\left(\bm{A}_1 \right)  \right]$ are the same for L1$_0$ and "40", graph energy of $\bm{A}_1$ for "40" is much higher than that for L1$_0$ as seen from Fig.~\ref{fig:cp}: This reflects the fact that $\mu_4 \left[ \mathrm{Spec}\left(\bm{A}_1 \right)  \right]$ for L1$_0$ is higher than that for "40", corresponding to the higher number of closed planar quartet links composed of 1NN pairs in L1$_0$ than that in "40".

\begin{figure}
\begin{center}
\includegraphics[width=0.91\linewidth]
{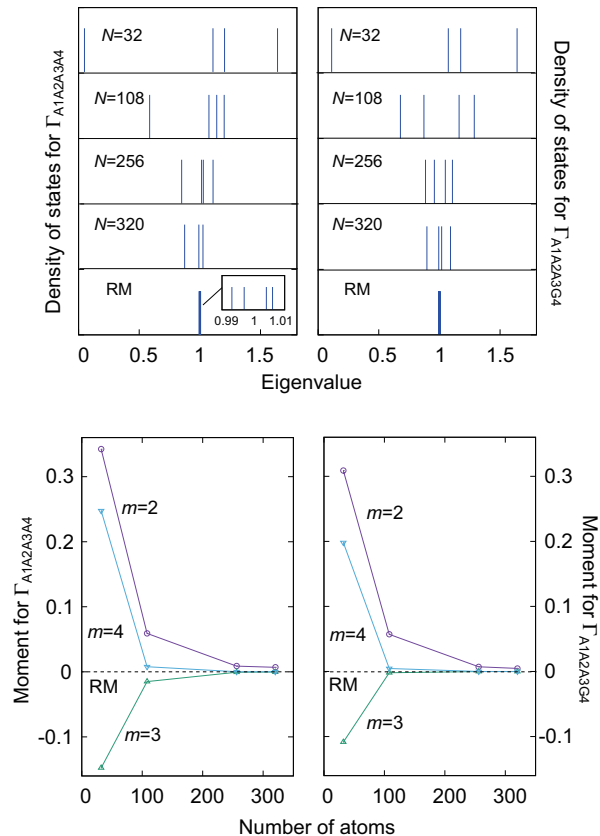}
\caption{(Color online) Upper: Density of states for eigenvalues of covariance matrices, $\Gamma_{A1A2A3A4}$ and $\Gamma_{A1A2A3G4}$ for multiple system size. Lower: Second, third and fourth order moment for the corresponding density of states as a function of number of atoms in the system.  }
\label{fig:doms}
\end{center}
\end{figure}

Another important investigation is the characteristics of density of microscopic states on configuration space (DOMS) for \textit{non-interacting} system.  
Note here that the non-interacting system itself appears not interesting, since practical system has interactions, and physical quantity should reflect such interactions. However, what we would like to investigate  in non-interacting system is different from this point. In our previous studies,\cite{emrs,lsi} we have found that physical quantity (especially, dynamical variables) for interacting classical system in equilibrium state, can be well characterized by physical quantity of a single special microscopic state whose structure can be known from information about non-interacting system: This certainly indicates that non-interacting system has significant information about how physical quantities in equilibrium state behaves when the interaction is provided to the system.  This fact relies on that DOMS for non-interacting system is well-characterized by multidimensional gaussian, which is further quantitatively studied based on random matrix: We confirm that when number of constituents increases, interdependence between structural degree of freedom numerically vanishes, which enables, for instance, direct estimation of free energy by single integration for analytic functions.\cite{yuge-RM,si} Based on law of large numbers, the former, i.e., DOMS  taking landscape of multidimensional gaussian, does not appear surprising, while the latter, the behavior of vanishment of interdependence for certain class of spatial constraint is non-trivial, since structural degree of freedom themselves do not independently take their values, certainly resulting in anisotropic landscape of configurational polyhedron shown in Fig.~\ref{fig:cp}. The present study focus on this non-trivial interdependence by considering whether our proposed graph representation exhibits similar interdependence behavior to GIM description previously considered. 

Therefore, we here consider DOMS for fcc conventional cell at equiatomic composition with different system size of 32, 108, 256 and 320 atoms, where individual atomic configuration is uniformly sampled based on Monte Carlo simulation with totally 102400 steps. To construct DOMS, we prepare two systems: One is four dimensional space of $\mathrm{Tr}\left[\bm{A}_i^2 \right]$ $\left(i=1,2,3,4\right)$, and another is also four dimensional space of $\mathrm{Tr}\left[\bm{A}_i^2 \right]$ $\left(i=1,2,3\right)$ and graph energy for $\bm{A}_4$. 
To see the statistical interdependence, we compare the density of eigenstates (DOE) of covariance matrix for the above four dimensional DOMS (hereinafter we describe as $\Gamma_{A1A2A3A4}$ and $\Gamma_{A1A2A3G4}$), with that for random matrix with gaussian orthogonal ensemble: Details for the present approach based on random matrix is described in our previous paper.\cite{yuge-RM} In brief, we numerically construct $4\times 102400$ random matrix and estimating corresponding covariance matrix and their eigenvalues. 
Figure~\ref{fig:doms} shows the resultant DOE and their $m$-th order moment ($m=2,3,4$), as a function of number of atoms in the system. 
We can clearly see that landscape of DOE for $\Gamma_{A1A2A3A4}$ qualitatively become close to that for random matrix when system size increases, which also holds for 
the system including information about graph energy, $\Gamma_{A1A2A3G4}$: The former tendency has been confirmed by our previous study.\cite{yuge-RM} 
This tendency can be quantitatively seen for the system size dependence of $m$-th order moment, where each moment gradually become close to that for random matrix with increase of system size. 
Note that DOE for random matrix does not approaches to delta function even when number of constituents goes to infinity, but exhibit finite variance.\cite{rm} We have previously found\cite{yuge-RM} that for a certain class of spatial constraint, the DOE on practical non-interacting system numerically gets closer rather to that for random matrix than to delta function: However, whether this behavior is universal for other class of constraints and its reason is still under discussion. Therefore, in the present study, we first see the similarity of asymptotic behavior of the interdependence between the proposed graph and GIM descriptions, shown in Fig.~\ref{fig:doms}. Although only from the figure, we do not conclude whether DOE for the proposed graph representation will coincide with that for random matrix when $N$ increases, they qualitatively show similar asymptotic behavior, indicating that our previously developed approach to determine equilibrium properties based on the information of spatial constraint can be extended to describing topological aspect of structures, by using the proposed representation of structures based on graph theory. Since we have confirmed that to see quantitative asymptotic behavior, we require much 
higher-dimensional configuration space, requiring high-computational cost especially for graph calculation, which should be further investigated in our future study.

\section{Conclusions}
We extend a concept of generalized Ising model (GIM) to graph theory, proposing a new representation to treat configuration-dependent physical quantities for crystalline solids: Landscape of resultant graph spectrum explicitly includes structure information in terms of both GIM and graph descriptions. We demonstrate the importance to consider linear combination of graph to find out additional characteristic structures compared with GIM, and to further investigate role of spatial constraint on equilibrium properties. We also address statistical interdependence for density of microscopic states using the proposed representation, exhibiting similar behavior that has been confirmed for the case of GIM description. This indicates that our previously developed approach to describe equilibrium properties based on the information of spatial constraint can be extended to including topological information of microscopic structures based on the proposed approach, which should be further addressed in our future study.


\begin{acknowledgements}
This work was supported by a Grant-in-Aid for Scientific Research (16K06704) from the MEXT of Japan, Research Grant from Hitachi Metals$\cdot$Materials Science Foundation, and Advanced Low Carbon Technology Research and Development Program of the Japan Science and Technology Agency (JST).
\end{acknowledgements}

\appendix
\section*{Appendix}
\begin{figure}[h]
\begin{center}
\includegraphics[width=0.50\linewidth]
{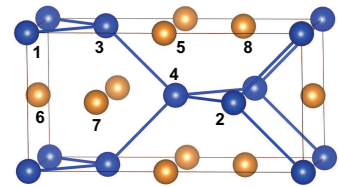}
\caption{(Color online) Atomic configuration of "40" structure found in the vertex of CP in Fig.~\ref{fig:cp} on $1\times 1\times 2$ expansion of fcc conventional cell. Dark blue spheres denote element with spin $\sigma=+1$, and bright orange ones denote that with spin $\sigma=0$. 1st nearest-neighbor pairs composed of the former element is described together. }
\label{fig:40}
\end{center}
\end{figure}

Here we show simple example of adjacency matrix for 1NN pair in binary system (with spin takes +1 and 0). Figure~\ref{fig:40} shows the atomic configuration of "40" ordered structure found in Fig.~\ref{fig:cp} based on expansion of fcc conventional cell. Then the corresponding $\bm{A}_1$ becomes
\begin{eqnarray}
\bm{A}_1= \left(
    \begin{array}{cccccccccccccccccccccccccccccccc}
0 & 1 & 1 & 0 & 0 & 0 & 0 & 0  \\
1 & 0 & 0 & 1 & 0 & 0 & 0 & 0  \\
1 & 0 & 0 & 1 & 0 & 0 & 0 & 0  \\
0 & 1 & 1 & 0 & 0 & 0 & 0 & 0  \\
0 & 0 & 0 & 0 & 0 & 0 & 0 & 0  \\
0 & 0 & 0 & 0 & 0 & 0 & 0 & 0  \\
0 & 0 & 0 & 0 & 0 & 0 & 0 & 0  \\
0 & 0 & 0 & 0 & 0 & 0 & 0 & 0  \\
     \end{array}
  \right),
\end{eqnarray}
where each row and column index corresponds to index of lattice point shown in Fig.~\ref{fig:40}. Each element having "1" comes from the product of basis function on each lattice point, i.e., 
$\sqrt{\phi_1\left(\sigma_i\right)\phi_1\left(\sigma_k\right)} = \sqrt{\sigma_i\sigma_k}=1$. 
To obtain the present result, for instance, of Fig.~\ref{fig:cp}, we use 256-atom cell (i.e., $4\times 4\times 2$ expansion of Fig.~\ref{fig:40}) under periodic boundary condition.

\end{document}